\documentstyle[12pt]{article}
\begin{document}
\tolerance=5000
\def\pp{{\, \mid \hskip -1.5mm =}}
\def\cL{{\cal L}}
\def\be{\begin{equation}}
\def\ee{\end{equation}}
\def\bea{\begin{eqnarray}}
\def\eea{\end{eqnarray}}
\def\tr{{\rm tr}\, }
\def\nn{\nonumber \\}
\def\e{{\rm e}}
\def\D{{D \hskip -3mm /\,}}
%%%%%%%%%%%%%%%%%%%%%%%%%%%%%%%%%%%%%%%%%%%%%%%%%%%%%%%%%55
\vfill

\begin{center}
{\Large\bf Inflationary Brans-Dicke Quantum Universe}

\vfill

{\sc Bodo Geyer$^{\diamondsuit}$}\footnote{
e-mail:geyer@ itp.uni-leipzig.de} 
, {\sc Sergei D. Odintsov$^{\spadesuit,\diamondsuit}$}\footnote{
e-mail: odintsov@mail.tomsknet.ru, odintsov@itp.uni-leipzig.de}
and 
{\sc  Sergio Zerbini }\footnote{
e-mail: zerbini@science.unitn.it }

\vfill

{\sl Department of  Physics and Gruppo Collegato INFN\\
University of Trento 38050-Trento, ITALY}

\

{\sl $\spadesuit$
Tomsk State Pedagogical University, 634041 Tomsk, RUSSIA}

\

{\sl $\diamondsuit$ 
Center of Theoretical Studies, 
 Leipzig University, Augustusplatz 10\\
04109 Leipzig, GERMANY}

\
\end{center}
%\vfill

\noindent
{\bf Abstract}
\\
\\
The formulation of Brans-Dicke (BD) gravity with matter in the Einstein 
frame is realized as Einstein gravity with dilaton and dilaton coupled 
matter. We calculate the one-loop 4d anomaly--induced effective action 
(EA) due to N dilaton coupled massless fermions on the time dependent 
conformally flat background with non-trivial dilaton. 
Using such (complete) effective action, the 
(fourth-order) quantum--corrected equations of motion are derived. 
One special
solution of these equations representing an inflationary Universe (with 
exponential scale factor) and (much slower) expanding BD dilaton is given. 
Similarly, 2d quantum BD Universe with time--dependent dilaton is constructed.
In the last case, the dynamics is completely due to quantum effects.

\vfill


\newpage

1. Brans-Dicke theory \cite{will} represents one of the simplest examples 
of scalar--tensor (or dilatonic) gravities where the background is described 
by the metric and the dilaton. There are many motivations to discuss such 
a theory.
First of all, the dilaton is an essential element of string theories and the
low--energy string effective action (for a recent review, see \cite{polch}) 
maybe considered as some kind of BD theory (with higher order terms).
Second, dimensional reduction of Kaluza-Klein theories may naturally lead 
to BD gravity. 
Third, dilatonic gravities are expected to have such important 
cosmological applications as in the case of (hyper)extended inflation
 \cite{la}. Note that recently there was some activity on
the study of BD cosmologies with a varying speed of light \cite{barrow}. 
Taking into account the renewed interest in scalar--tensor gravity 
(see ref.\cite{FGT} for a review) it may be reasonable to 
investigate the quantum aspects of such theory. That will be the
purpose of present Letter: We apply effective action formalism 
for the dilaton coupled matter and study quantum cosmology 
in matter--BD gravity.

2. Let us start from the standard Brans-Dicke 4-dimensional action in the 
Jordan frame:
\be
S_{BD}=\frac{1}{16\pi}\int d^4x \sqrt{-g} \left[ 
\phi R -\frac{\omega}{\phi}
(\nabla_\mu \phi)(\nabla^\mu \phi) \right]+S_M\,,
\label{bdj}
\ee
where $\phi$ is the Brans-Dicke (dilaton) field with
$\omega$ being the coupling constant and $S_M$ is the matter action. 

It has been argued (for a review, see \cite{FGT}) that only the action in the
Einstein frame is physically relevant. For this reason, we prefer working 
within this frame, performing the following conformal transformation
\be
\tilde{g}_{\mu \nu}=G\phi g_{\mu \nu}
\ee
and a redefinition of the scalar field
\be
\tilde{\phi}=\sqrt{{2\omega+3}/(16\pi)} \ln \phi\,,\,\,\,\,
2\omega+3 >0\,. 
\ee
Thus, the action in the Einstein frame  reads
\be
S=\int d^4x \sqrt{-\tilde{g}(x)} \left[\frac{\tilde{R}}{16\pi G}
-\frac{1}{2}
(\tilde{\nabla}_\mu \tilde{\phi})(\tilde{\nabla}^\mu \tilde{\phi})+
\exp{(A\tilde{\phi})}L_M(\tilde{g}) \right]
\,,
\label{bde}
\ee
where $A=-8\sqrt{\frac{\pi G}{2\omega+3}}$
(for a recent review of dilaton status, see \cite{Dick}).

Below, we consider the theory in the  Einstein frame  as physical 
theory, thus we omit the tildes in order to avoid confusion. As matter
Lagrangian we take the one associated with $N$ massless (Dirac) spinors, i.e. 
\be
\label{matterOA}
L_M=\sum_{i=1}^N \bar\psi_i\gamma^\mu\nabla_\mu\psi^i\ .
\label{sd}
\ee
There is no problem to add other types of matter 
(say scalar or vector fields).
The choice (\ref{sd}) is made only for the sake of simplicity.

The action (\ref{bde}) describes the Einstein theory with a scalar (dilaton)
field and dilaton coupled spinor matter. Without matter, it may be also 
considered as a low--energy string effective action (EA) (even in the 
presence of some dilaton coupled matter).

Our purpose here will be to study the role of quantum effects associated with
 dilaton coupled spinors to cosmological problems in BD gravity. 
Assuming $N$ sufficiently 
large (to allow for the large  $N$ approximation) one can neglect 
in such investigation the  
 proper quantum gravity corrections. We shall make use of the 
EA formalism (for an introduction, see \cite{BOS}). 
The corresponding 4d anomaly--induced EA for dilaton coupled
scalars, vectors and spinors has been found in Refs. \cite{NO,NOS,NNO} 
respectively.

Hence, starting from the theory with the action
\be
S=\int d^4x \sqrt{-g} \left[\frac{R}{16\pi G}
-\frac{1}{2}
(\nabla_\mu \phi)(\nabla^\mu \phi)+
\exp{(A\phi)}\sum_{i=1}^N \bar\psi_i\gamma^\mu\nabla_\mu\psi^i  \right]
\,,
\label{bde1}
\ee
we will discuss FRW type cosmologies
\be
ds^2=-dt^2+a(t)^2 dl^2\,,
\label{st4}
\ee
where $dl^2$ is the line metric element of a 3-dimensional space with 
constant curvature $\Sigma$, namely $k=1$ ($\Sigma=S^3$), $k=0$  
($\Sigma=R^3$) or $k=-1$ ($\Sigma=H^3$).
As is well known, introducing the conformal time $\eta$ by means of
\be
dt=a(\eta) d\eta\,,
\ee
one gets a  space--time which is conformally related to an ultrastatic 
space--time with constant curvature spatial section $\overline{M}$, namely
\be
ds^2=a(\eta)^2 (-d\eta^2+ dl^2)=a(\eta)^2  d\bar{s}^2\,,
\label{stc}
\ee
or
\be
g_{\mu \nu}=e^{2\sigma(\eta)} \bar{g}_{\mu \nu}\,,
\ee
with
\be
a(\eta)=e^{\sigma(\eta)}\,.
\ee

The computation of the anomaly--induced EA for the dilaton coupled spinor 
field has been done in \cite{NNO}, and
the result, in the non-covariant local form, reads: 
\bea
\label{vii}
\lefteqn{\hspace{-.8cm}
W=\int d^4x \sqrt{-\bar g} \bigg\{b \bar F \sigma_1
+ 2b' \sigma_1\Big[ \bar{\Box}^2
+ 2 \bar R^{\mu\nu}\bar\nabla_\mu\bar\nabla_\nu 
%\right. \nn && \left. 
- {2 \over 3}\bar R\bar{\Box}
+ {1 \over 3}(\bar\nabla^\mu\bar R)\bar\nabla_\mu
\Big]\sigma_1}
\nn
&&\!\!\!\!\!\!\! %\qquad
+\, b'\sigma_1\Big(\bar G -{2 \over 3}\bar\Box\bar R\Big) 
%\nn && 
-{1 \over 18}(b + b')\left[\bar R - 6 \bar{\Box} \sigma_1
- 6(\bar\nabla_\mu \sigma_1)(\bar\nabla^\mu \sigma_1)
\right]^2\bigg\}\,,
\eea
where $\sigma_1=\sigma+ A\phi /3$, 
the square of the Weyl tensor is given by
$ F= R_{\mu\nu\rho\sigma}R^{\mu\nu\rho\sigma}
-2 R_{\mu\nu}R^{\mu\nu} + {1 \over 3}R^2 $
and Gauss-Bonnet invariant is
$G=R_{\mu\nu\rho\sigma}R^{\mu\nu\rho\sigma}
-4 R_{\mu\nu}R^{\mu\nu} + R^2$.
For Dirac spinors
$b={3N \over 60(4\pi)^2}$, $b'=-{11 N \over 360 (4\pi)^2}$.

Generally speaking, it should be noted that the complete 
one-loop EA is given by the anomaly--induced EA presented above, plus some
conformally invariant functional which is the EA computed in the reference
metric $\bar g_{\mu \nu}$. In our case, this second term is rather trivial
(actually, it is a $k$--dependent constant),
being the EA of a free spinor field in an ultrastatic space--time with 
constant spatial section (for a discussion of such effective actions, see 
\cite{Zerbini}). 

For example, for the flat case ($k=0$), 
$\bar g_{\mu \nu}=\eta_{\mu \nu}$, the Minskowski tensor. As a consequence,
$W$ gives the complete non-trivial EA!

For the sake of simplicity, let us consider the case $k=0$. As a result,
the EA becomes
 \be
\label{1ef}
W=V_3\int d\eta \left\{2b' \sigma_1 \sigma_1''''
- 2(b + b')\left( \sigma_1'' - {\sigma_1'}^2 \right)^2\right\}\ .
\label{7}
\ee
Here, $V_3$ is the (infinite) volume of 3-dimensional flat space and
$'\equiv{d / d\eta}$ and $\sigma=\ln a$ where
$a(\eta)$ is the scale factor.

The total one--loop EA may be written adding to $W$ the classical action
for $k=0$: 
\be
S=V_3\int d\eta \left[ \frac{6}{16 \pi G}(\sigma''+{\sigma'}^2) 
e^{2\sigma}+\frac{1}{2} {\phi'}^2 e^{2\sigma}\right]\,.
\label{ca}
\ee
Note that the notations are the same as in Ref.~\cite{B} 
where similar dilatonic effects to quantum cosmology
have been considered for $N=4$ super Yang Mills theory.
In that case, the dilaton appears as part of the conformal 
supergravity multiplet (classical background) and 
the corresponding terms in EA have a different form.

Let us write the equations of motion
\bea
&&\tilde{C} e^{({A \phi}/{3})}+\frac{12}{16\pi G}a''+a \phi'^2=0 \nn 
&& \frac{A}{3}\tilde{C} a e^{({A \phi}/{3})}-(a^2 \phi')'=0\,,
\label{8}
\eea
where
\bea
\tilde{C}&=&
-\frac{4b}{\tilde{a}} \left[ \frac{\tilde{a}''''}{\tilde{a}}
-\frac{4\tilde{a}' \tilde{a}'''}{\tilde{a}^2} 
- \frac{3\tilde{a}''^2}{\tilde{a}^2}\right] 
-\frac{24}{\tilde{a}^4} \left[ (b-b') \tilde{a}'' \tilde{a}'^2 
+ b'\frac{\tilde{a}'^4}{\tilde{a}} \right] \, \nn
&=&
-\frac{4b}{\tilde{a}} 
\bigg[ \bigg(\frac{\tilde{a}'}{\tilde{a}}\bigg)^{\prime\prime}
-2\Big(1+\frac{b'}{b}\Big) \bigg(\frac{\tilde{a}'}{\tilde{a}}\bigg)^{3} 
\, \bigg]^{\prime} 
\label{88}
\eea
\be
\tilde{a}=a  e^{({A \phi}/{3})}\equiv a(\eta) e^{({A \phi(\eta)}/{3})}
\,. 
\ee
Combining Eqs.~(\ref{8}), one gets
\be
a \frac{A}{3} \left[\frac{12}{16\pi G}a''+a \phi'^2 \right]+
(a^2 \phi')'=0\,.
\label{9}
\ee
Of course, it is hopeless to find the general solution of Eqs. (\ref{8}).
However, we remind that there exists a  well known solution in the absence
of the dilaton,
namely Einstein gravity with quantum matter. Such solution describes the
inflationary Universe (see \cite{star}); for a recent discussion
of non--singular string cosmologies, see \cite{Veneziano}.
 Having in mind such inflationary 
Universe solution, one can search for special solution of (\ref{9}), with
\be
a(\eta)=\frac{1}{H \eta}\,,\,\,\,\,\,\, 
\phi'(\eta)=\frac{1}{H_1 \eta}\,,
\label{10}
\ee
where $H$ and $H_1$ are some constants. Substituting Eq.~(\ref{10}) in 
(\ref{9}), we obtain
\be
\frac{3}{2 \pi G} H^2_1-\frac{9}{A}H_1+1=0\,. 
\label{11}
\ee
Hence, time--dependence and $H$--dependence decouple and one is 
left with a second order algebraic equation for $H_1$. Its solution is
\be
H_1=\sqrt{4\pi G}\left[- \frac{3}{16}\sqrt{2\omega+3}\pm 
\sqrt{\frac{81}{4}(2\omega+3)-\frac{1}{6}} \right]\,.
\label{12}
\ee
 From this equation one gets a quantum estimate for 
$\omega$: $\omega=-\frac{3}{2}+x$, $x > 3^{-5}$ 
that is slightly bigger than the classical restriction  $x>0$. 
Notice that $H_1$ is always negative for the minus sign, while for the 
plus sign it may be positive or negative depending on the value 
of BD parameter $\omega$.

Before analyzing Eq.~(\ref{8}) with the Ansatz (\ref{10}), let us check 
 that for $A=0$ and $\phi=\mbox{const}$ one  obtains the inflationary 
Universe solution. Indeed, using $a(\eta)={1}/{H\eta}$ in (\ref{8}), 
it is easy to get
\be
H^2=-\frac{1}{16\pi G b'}\,.
\label{13}
\ee  

Consider now the case of non-trivial dilaton $\phi = \phi(\eta)$.
Transforming back to the physical time $t$, one obtains the 
inflationary universe \cite{star}. Indeed, from
$dt=a(\eta)d\eta$ and $a(\eta)=\pm {1}/{H\eta}$, one obtains 
 $a(t)=1/C\exp{(\mp Ht)}$. 
Choosing the plus sign gives
rise to the inflationary universe solution. In this case, $\phi \sim
\frac{H}{H_1}t$, namely the dilaton is expanding with time, but much slower
than the scale factor.

Let us now use the Ansatz (\ref{10}) and Eq.  (\ref{12}) in the first of the 
Eqs. (\ref{8}), in order to find $H^2$. To this aim, first we get
\be
\tilde{a}(\eta)=\frac{1}{H}\eta^{\frac{A}{3H_1}-1}\,.
\label{14}
\ee  
Notice that according to estimations using Solar System experiments 
\cite{will}, $\omega >500$. As a consequence,
for $\omega=500$ one gets $\frac{A}{3H_1}\simeq \pm 
0.0003$ (depending on the sign choice for $H_1$). Hence, $\frac{A}{3H_1}$
can be neglected in the above Equation.

Thus, from the first of Eqs. (\ref{8}), one has
\be
H^2=-\frac{1}{16\pi G b'}-\frac{1}{24H_1^2 b'}\,.
\label{15}
\ee
As a result, one observes that the value of $H^2$ is 
significally increased by
the contribution associated with the dilaton!

Summarizing, we have found explicitly one special analytic solution
describing Brans-Dicke non-singular
 Universe with a (much slower) expanding dilaton. 
It should be noted that this is a purely quantum solution which does not 
exist at classical level. Note also that when $|\frac{A}{3H_1}|$ is very 
large, $\omega\simeq 3/2$ one can easily get another estimatations 
for $H^2$ which depends now on the explicit choice of 
BD parameter $\omega$. It comes mainly from dilaton.
 
3. Motivated by the 4-dimensional case, we may assume that the classical 
action in the Einstein frame for the 2-dimensional BD theory reads
\be
S=\int d^2x \sqrt{-g(x)} \left[ \frac{R}{16\pi G}-\frac{1}{2}
(\nabla_\mu \phi)(\nabla^\mu \phi)+
\exp{(A \phi)L_M} \right]\,,
\label{s2}
\ee
where $A$ is a constant parameter of the 2d BD theory and the matter 
Lagrangian is the one of two-dimensional Majorana spinors:
\be
\label{i}
L=\sum_{i=1}^N \bar\psi_i\gamma^\mu\nabla_\mu\psi^i\ .
\ee
Let us neglect the classical matter contribution since we are interested 
only in the one-loop EA induced by the conformal anomaly of the quantum 
matter. 

In two dimensions, the metric of the Universe  may be written in  the form
\be
ds^2=-dt^2+a(t)^2 dx^2\,.
\label{st}
\ee
Introducing again the conformal time 
one gets a conformally flat space-time, 
\be
g_{\mu \nu}=e^{2\sigma(\eta)}\eta_{\mu \nu}\,.
\ee
We also assume that the BD scalar field  $\phi$ depends only on 
$\eta$.

Since the conformal anomaly for the dilaton coupled spinor
 is  (see \cite{NNO})
\be
T=c\left[{1 \over 2}R + 2\triangle \phi \right]\ ,
\ee
where $c={N /( 12\pi)}$, 
the anomaly induced EA  in the local, non-covariant form  reads 
\be
\label{eas}
W=c\int d^2x 
\left\{-{1 \over 2}\sigma \sigma'' -
A\sigma \phi'' \right\}\ .
\ee
%Here $'=\frac{d}{d\eta}$.
Note that this is, up to a
non--essential constant, an exact expression.  
The total one--loop effective action is $S+W$, i.e.
\be
\label{tot}
S+W=
V_1\int d\eta  \left\{-\frac{k_G}{2} \sigma''+\frac{1}{2} (\phi')^2 
-\frac{c}{2}\sigma \sigma''-cA \sigma \phi''  \right\}\ \,,
\ee
where $k_G=\frac{1}{4 \pi G}$ and  $V_1$ the (infinite) spatial volume.
After integration by parts we obtain %w.r.t. $d\eta$
\be
\label{tott}
S+W=
V_1\int d\eta \left\{ \frac{1}{2} 
{\phi'}^2+cA \phi' \sigma'+\frac{c}{2} \sigma'^2 \right\}\,.
\ee
As we see, 2d Einstein theory is trivial. It has no dynamics 
associated with scale factor (but th edilaton has some dynamics 
from classical part). The whole dynamics (or cosmology) 
appears as a result of quantum effects.

The general solution  for $a(t)$ is a linear function of time 
plus a constant:
\be
a(t)=a_0+a_1t\,.
\ee
For the dilaton we get 
\be
\phi(t)=\phi_0+\Big(\frac{h}{a_1}-cA\Big)\ln a(t)\,.
\ee
In this case, the initial singularity at $t=0$ seems to be absent
(however, curvature is zero at all times). Note in this respect 
that previous attempts\cite{cosm} to study quantum cosmology in 2d dilatonic gravity
have been actually studied in the Jordan frame.
 
In summary, we presented the first consistent framework 
to study quantum BD gravity with matter in the Einstein frame.
The main problem here was to find the effective action for dilaton
coupled matter. We successfully constructed such effective action
for conformally flat background with non--trivial dilaton 
(only the spinor case was discussed). That gave the possibility 
to investigate quantum BD cosmology.
One 4d analytic solution found in this Letter describes
non--singular (inflationary) Universe with exponentially growing 
scale factor and linearly growing dilaton. In 2d case dynamics is induced
 only by quantum effects and non-trivial cosmology maybe also constructed.

It would be very interesting to extend the results of this study 
in various directions.
First of all, we searched for an explicit analytic solution of 
the equations of motion.
Clearly, there exist many more solutions describing different 
types of BD cosmology (expanding, oscillating, etc). 
These solutions may be found 
numerically for different initial conditions chosen for the dilaton and
the scale factor, and also for their derivatives. Another extension
is related with the possibility to study in detail $k=1$ 
or $k=-1$ quantum Universes as the corresponding effective action
is derived in this work. These questions will be addressed 
in another place.
 
\noindent
{\bf Acknowledgments}.
The research by
SDO was partially supported by a RFBR Grant N\,99-02-16617,
by Saxonian Ministry of Science and Arts and by Graduate College
``Quantum Field Theory" at Leipzig University.

\end{document}